\documentstyle[12pt,psfig,myplots,aaspp]{article}
\begin{document}

\textwidth 6.5in
\textheight 8.5in
\topmargin 0.0in
\oddsidemargin 0.0in

\title{Lunar Outgassing, Transient Phenomena and The Return to The Moon\\
III: Observational and Experimental Techniques}

\author{Arlin P.S.~Crotts}
\affil{Department of Astronomy, Columbia University,
Columbia Astrophysics Laboratory,\\
550 West 120th Street, New York, NY 10027}

\begin{abstract}
In Paper I of this series, we show that transient lunar phenomena (TLPs)
correlate with lunar outgassing, geographically, based on surface radon release
episodes versus the visual record of telescopic observers (the later prone to
major systematic biases of unspecified nature, which we were able to constrain
in Paper I).
In Paper II we calculate some of the basic predictions that this insight
implies, in terms of outgassing/regolith interactions.
In this paper we propose a path forward, in which current and forthcoming
technology provide a more controlled and sensitive probe of lunar outgassing.
Many of these techniques are currently being realized for the first time.

Given the optical transient/outgassing connection, progress can be made by
Earth-based remote sensing, and we suggest several programs of imaging, 
spectroscopy and combinations thereof.
However, as found in Paper II, many aspects of lunar outgassing seem likely to
be covert in nature.
TLPs betray some outgassing, but not all outgassing produces TLPs.
Some outgassing may never appear at the surface, but remain trapped in the
regolith.

As well as passive remote sensing, we also suggest more intrusive
techniques, from radar mapping to in-situ probes.
Understanding these volatiles seems promising in terms of their exploitation as
a resource for human presence on the Moon and beyond, and offers an
interesting scientific goal in its own right.

This paper reads, therefore, as a series of proposed techniques, some in
practice, some which might be soon, and some requiring significant future
investment (some of which may prove unwise pending results from predecessor
investigations).
These point towards enhancement of our knowledge of lunar outgassing, its
relation to other lunar processes, and an increase in our understanding of how
volatiles are involved in the evolution of the Moon.
We are compelled to emphasize certain ground-based observations in time for
the flight of {\it SELENE, LRO} and other robotic missions, and others before
extensive human exploration.
We discuss how study of the lunar atmosphere in its pristine state is pertinent
to understanding the role of anthropogenic volatiles, at times a significant
confusing signal.
\end{abstract}

\medskip
\section{Introduction}

Transient lunar phenomena are defined for the purposes of this investigation as
localized (smaller than a few hundred km across), transient (up to a few hours
duration, and probably longer than typical impact events - less than 1s to a
few seconds), and presumably confined to processes near the lunar surface.
How such events are manifest is summarized by Cameron (1972).
In Paper I we study the systematic behavior (especially the spatial
distribution) of TLP observations - particularly their significant correlations
with tracers of lunar surface outgassing, and in Paper II some simple,
theoretical predictions of other, not-so-obvious aspects that might be
associated with TLPs and outgassing events.
In this paper we suggest several ways that more information might be gleaned to
determine the true nature of these events.
At several points we emphasize the importance of timely implementation of these
approaches.

TLPs are infrequent and short-lived, and this is the overwhelming fact of their
study that must be surmounted.
It is our goal to design a nested system of observations which overcomes the
problems that this fact has produced, a largely anecdotal and bias-ridden data
set, and replace it with another data set with {\it a priori} explicit,
calculable selection effects.
This might seem a daunting task, since the data set we used in Paper I was
essentially the recorded visual observations of the entire human race since the
invention of the telescope, and even somewhat before.
With modern imaging and computer technology, however, we can overcome this.

Another problem that becomes clear in Paper II is the many, complex means by
which outgassing can interact with the regolith.
In the case of slow seepage, gases may take a long time to work their way
through the regolith.
If the gases are volcanic, there may be interactions along the way, and if
water vapor is involved, it and perhaps others of these gases may remain
trapped in the regolith.
These factors must be remembered in designing our future investigations.

We can make significant headway, however.
The various factors which complicate our task due to the paucity of information
about TLPs also leave open avenues that modern technology can exploit.
The many methods detailed in this paper are summarized in Table 1.
There has been no areal-encompassing, digital image monitoring of the near side
with appreciable time coverage using modern software techniques to isolate
transients.
There are no published panspectral maps at high spectral/spatial resolution of
the near side surface, beyond what is usually called multispectral imaging.
(To some degree this will be achieved by the Moon Mineralogy Mapper onboard
{\it Chandrayaan-1}, but not before other relevant missions such as $SELENE$
have passed).
There are numerous particle detection methods that are of use.
The relevant experiments on Apollo were of limited duration, either of a week
or less, or 5-8 years in the case of ALSEP.
Furthermore the {\it Clementine} and {\it Lunar Prospector} missions were also
of relatively short duration.
All of these limitations serve as background to the following discussions.

\section{Optical/Infrared Remote Sensing}

\subsection{Earth-Based Imaging}

By necessity the monitoring of optical transients from the vicinity of Earth
must be limited to the near side.
As detailed in Paper I, however, all physical correlations tied to TLPs
likewise strongly favor the near side e.g., $^{222}$Rn outgassing (4 of 4
episodes being nearside, as well as nearly all $^{222}$Rn residual (seen as
$^{210}$Po) and mare edges ($\sim$85\% nearside, somewhat depending on one's
definition, even more so if low-contrast albedo features such as Aitken basin
are not included).

Remote sensing in the optical/IR is limited in spatial resolution either by the
diffraction limit of the telescope or by atmospheric seeing.
One arcsecond, a typical value for optical imaging seeing FWHM, corresponds to
1.8-2.0~km on the lunar surface, and is the diffraction limit of a 12~cm
diameter telescope at $\lambda = 600$~nm.

The best, consistent imaging resolution will come from the $Hubble$ $Space$
$Telescope$ with $0.07-0.1$ arcsec FWHM, and indeed images of the Moon have
been obtained with the $HST$/Advanced Camera for Surveys combination (Garvin
et al.~2005).
$HST$ observations of the Moon turn out to be relatively expensive in terms of
spacecraft time due to setup time complicated by the relative motions of the
target and spacecraft, and inefficiency due to exposure setup times of
$\sim$80s for each exposure of typically 1s.
Altogether $\sim 0.5$-1~h of spacecraft time is needed to successfully image a
small region in one filter band (due in part to several overlapping exposures
needed for complete coverage avoiding masks and other obstructions on the HRC
detector, as well as to reject cosmic ray signals).
At least until the $Hubble$ Servicing Mission 4, the guiding of $HST$ and the
state of ACS will allow no further such observations.

A competing method for producing high-resolution imaging is the ``Lucky
Exposures'' (LE, also ``Lucky Imaging'') technique which exploits occasionally
superlative imaging quality among a series of rapid exposures, then sums the
best of these with a simple shift-and-add algorithm (Fried 1978, Tubbs 2003).
The technique requires a high-speed, linear-response imager, and can be
accomplished only with great difficulty using a more conventional astronomical
CCD system.
Nonetheless, many amateur setups have achieved excellent results with this
technique, and the Cambridge group (Law, Mackay \& Baldwin 2006) have achieved
diffraction-limited imaging on a 2.5-meter telescope, very close to $HST$
angular resolution.
In practice, only about 1-10\% of exposures, hence less than 1\% of observing
time, survive image quality selection, but for the Moon this amounts to a small
investment of telescope time (a few minutes). 
We have attempted this ourselves and encountered some minor problems: image
quality must be selected in terms of a fourier decomposition of the image
rather than inspection of the point-spread function of a reference star, and
shift-and-add parameters must be similarly defined, by image cross-correlation
rather than by centroiding a bright star.
We will present results from these efforts when they succeed more usefully.

Unlike adaptive optics approaches, LE does not depend on a bright reference
star to define the incoming wavefront, but LE improvements are still limited to
an angular area of the isoplanatic patch determined by atmospheric turbulence,
$\sim$1000 arcsec$^2$.
Covering the entire nearside Moon would be challenging ($\sim$3000 fields
needed - at least 20 nights on a moderate-sized telescope).
Likewise, the ACS HRC on $HST$, covering 750 arcsec$^2$ at a time, cannot be
used practically to map the entire near side.
The greater flexibility of an LE program, in terms of choice of epoch and
wavelength coverage, provides many advantages; ACS HRC, on the other hand,
would provide consistent-quality results, albeit at great expense.

High resolution imaging can be used to monitor small, specific areas over time,
or in a one-shot application comparing a few exposures to imaging from another
source.
Currently, the best full-surface comparison map in the optical is the 
$Clementine$ UVVIS CCD map (Eliason et al.~1999), 5 bands at 415-1000~nm, with
typically 200m resolution, a good match to LE and $HST$ resolutions.
Unfortunately, neither $Clementine$ UVVIS, or infrared cameras NIR (1100-2800
nm) or LWIR (8000-9500 nm) cover some of the more interesting bands for our
purposes (for example, the regolith hydration bands at 2.9 and 3.4 $\mu$m).
In the future, we will be able to make comparisons to the extensive map of the
Moon Mineralogy Mapper (Pieters et al.\ 2005) on Chandrayaan-1, with 140~m
and 20~nm FWHM spatial and wavelength resolution, respectively, over
0.4-3.0~$\mu$m.

The 3 $\mu$m-reflectance hydration features in asteroidal regolith have been
studied (Lebofsky et al.\ 1981, Rivkin et al.\ 1995, 2002, Volquardsen et
al.\ 2004).
There is little written about the spectroscopic reaction of lunar regolith to
hydration; however, it is apparent that the reflectance features near 3 $\mu$m
do not appear immediately in lunar samples subjected to the terrestrial
atmosphere (Akhmanova et al.~1972), but do after several years (Markov et
al.~1980, Pieters et al.\ 2006).
At least in the latter, samples lose this hydration reflectance effect within a
few days of exposure to a dry environment.
This issue could easily be studied with further lunar sample experiments.

The prime technique for detecting changes between different epochs in similar
images will involve image subtraction.
This technique is well-established in studying supernovae, microlensing and
variable stars, and produces photon Poisson noise-limited performance (Tomaney
\& Crotts 1996).
This technique is well matched to CCD or CMOS detectors, and at 1-2 arcsec
FWHM resolution, these can cover the whole Moon with 10-20 Mpixels, as is
available for conventional detectors.
For proper image subtraction, one needs at least 2 pixels per FWHM diameter,
or else non-Poisson residuals tend to dominate, driving up the variable source
detection threshold.

To illustrate how image subtraction would work, we present data of the kind
that might be produced by a monitor to detect TLPs.
While the image shown in Figure 1 is taken on a 0.9-meter telescope with
24~$\mu$m pixels, the data are similar to that would be produced by a smaller,
1-arcsec diffraction-limited telescope with typical commercially-available
digital-camera pixels e.g., 6~$\mu$m on a 20-cm telescope.

Image subtraction delivers nearly photon-noise level accuracy in the residual
images taken in a ground-based time series, and this is demonstrated in Figures
2-4.
We introduce an artificial ``TLP'' signal that is a 8\% enhancement over the
background in the peak pixel of an unresolved source - a signal at or below the
threshold of a visual search.
The TLP is detected convincingly even in a single image, once subtracted from
a reference image e.g., the average of a time series.
The subtraction gives a very flat residual subtracted image (except for the
simulated TLP and a few ``cosmic rays'' of much smaller area and amplitude).
The only exception is in the complex image region of the highlands near the
global terminator.

More meaningful, perhaps, is the signal-to-noise ratio of residual sources,
shown in Figure 3.
This shows the TLP clearly and unambiguously, but there are some false
detections in the highland local terminator region at the level of 10-20\% of
the TLP; we would like to improve on this.
One alternative to reduce this noise is to consider applying an edge filter to
supply a weighting function to suppress regions where the image structure is
too complex.
Figure 3 shows the result from processing the raw image with a Roberts edge
enhancement filter ($G_{j,k}=|F_{j,k} - F_{j+1,k+1}|+|F_{j,k+1} - F_{j+1,k}|$,
where $F_{j,k}$ is the raw count in the pixel $(j,k)$ and $G$ is the function
shown in Figure 3).
When the signal difference from Figure 2 is divided by Figure 3, the result
(Figure 4) uniquely and clearly shows the TLP.
We would like to avoid this edge filter strategy if possible, relying
completely on simple image subtraction, since it may be that some TLPs are
associated with local terminators on the lunar surface.

Our group has automated a TLP monitor on the summit of Cerro Tololo that
should be producing regular lunar imaging data as of mid-2007 (Crotts,
Hickson \& Pfrommer 2007).
This will cover the entire Moon at 1 arcsec resolution, and we expect to be
able to process the images at a rate of one per 10s.
This is sufficient to time-sample nearly all reported TLPs (see Paper I).
In addition we plan to add a second imaging channel on a video loop; this will
retain a continuous record of imaging of sufficient duration so that an alert
to a TLP event from the image subtraction processing pipeline will allow one to
query the image cache of the video channel record and reconstruct the event at
finer time resolution.
The image subtraction channel will include a neutral-density filter to allow
the exposure time to nearly equal the image cycle time, hence even short TLPs
(or meteorite impacts) will be detected, albeit at a sensitivity reduced by a
factor roughly proportional to the square-root of the event duration.

The presence of a lunar imaging monitor opens many possibilities for TLP
studies.
For the first time, this will produce an extensive, objective, digital record
of changes in the appearance of the Moon, at a sensitivity level much finer
than the capability of the human eye.
While we will see the true frequency of TLPs soon enough, Paper I indicates
that perhaps one TLP per month might be visible to a human observer observing
at full duty cycle.
An automated system should be able to distinguish changes in contrast at the
level of 1\% or slightly better, whereas this is perhaps 10\% for a point
source observed by the human eye (based on our tests).
Even augmented human-eye surveys (such as Project Moon Blink or the Corralitos
Observatory TLP survey - see Paper I) would be at least several times less
sensitive than a purely digital survey.
The resulting frequency of TLP detections at higher sensitivity depends on the
event luminosity distribution function, poorly defined even at brighter limits
and completely unknown at the level that will now be accessible.
It might be reasonable to assume that a single monitor might detect several
TLPs per month of observing time.
Over several years, monitors at a range of terrestrial longitudes might detect
of order 100 or more TLPs, providing a well-characterized sample that will
avoid many of the selection problems of the anecdotal visual data base and
approach similar sample sizes.

Our plan eventually is to run two or more such monitors independently.
Not only does this increase the likely TLP detection rate, but allows us to
perform simultaneous imaging in different bands, or in different polarization
states.
Dollfus (2000) details TLPs evident as polarimetric anomalies.
The timescales involved are not tightly constrained, between 6~min and 1~d.
Other transient polarimetric events (Dzhapiashvili \& Ksanfomaliti 1962, Lipsky
\& Pospergelis 1966) are even less constrained temporally; however, the fact
that we can observe the same event with two monitors simultaneously (while
observing the rest of the Moon), means that there is little systematic doubt
concerning the degree of polarization due to variability of the source while
the apparatus is switching polarizations.
Presumably, since these are likely due to simple scattering effects on linear
polarization, we should align the E-vector of one monitor's polarizer parallel
to the Sun-Moon direction on the sky, and the second perpendicular to it.
In the case of three or four monitors operating simultaneously, we can
reconstruct Stokes parameters for linear polarization conventionally by
orienting polarizer E-vectors every $60^\circ$ or $45^\circ$, respectively.
The total flux from two or more monitors can be obtained by summing in
quadrature signals from the different polarizations.

A TLP imaging monitor will also open new potential as an alert system for other
observing modes.
A monitor detection can trigger LE imaging in a specific active area.
A qualitatively unique possibility is using the monitor to initiate 
spectroscopic observations, which much better than imaging will provide
information about non-thermal processes and perhaps betray the gas associated
with the TLP.

\subsection{Ground-Based Spectroscopy/Hyperspectral Observations}

TLP spectroscopy has its challenges.
In order to detect a change, we must make comparisons over a time series of
spectroscopic observations.
This is essentially a four-dimensional independent-variable problem, therefore:
two spatial dimensions of the lunar surface, plus wavelength implying a data
cube, plus time.
Whereas ``hyperspectral'' imaging usually refers to a resolving power
$R = \lambda / \Delta \lambda \approx 50-100$, where $\Delta \lambda$ is the
FWHM wavelength resolution, the emission lines from TLPs might conceivably be
many times more narrow than this, thereby diluted if higher resolution is not
employed.
It is not currently conceivable to monitor the whole near side in this way (at
$\sim 1$ Gpixel s$^{-1}$ for $R=1000$ and an exposure every 10s), but this is
unnecessary.
A practical approach may be to set up the reduction pipeline of the TLP monitor
to alert to an event during its duration e.g., in under 1000s, and then to
bring a larger telescope with an optical or IR spectrograph to bear on the
target, which our experience shows might be accomplished in $\sim$300s.
We are working to implement this in 2007.

There are reasons to prepare an $R \approx 300$ data cube in advance of a TLP
campaign for reasons beyond simply having a ``before'' image of the Moon prior
to an event.
For instance, in the IR there are regolith hydration bands near 2.9 and
3.4~$\mu$m, the latter with substructure on the scale of $\sim$20~nm, which
will be degraded unless the instrumental resolution is $R \ga 300$.
While there are fewer narrow features in the optical/near-IR, the surface
Fe$^{2+}$ feature at 950~nm of pyroxene (which requires only $R \approx 10$ to
be resolved), shows compositional shifts in wavelength centroid and width on
the scale of $\sim$10~nm (Hazen, Bell \& Mao 1978), which requires $R \approx
100$ to be studied in full detail.
Likewise, differentiating pyroxenes from iron-bearing glass (Farr et al.~1980)
requires $R \approx 50$.
This Fe$^{2+}$ band (and the corresponding band near 1.9$\mu$m) are useful for
lunar surface age-determination since they involve surface states that are
degraded by micrometeorites and solar wind in agglutinate formation (Adams
1974, Charette et al.\ 1976).
It appears that overturn of fresh material can also be monitored with enhanced
blue optical broadband reflectivity (Buratti et al.\ 2000).

Such datasets are straightforward to collect, as are their reduction (although
requiring of some explanation).
Observations involve scanning across the face of the Moon with a long slit
spectrograph, which greatly improves the contrast of an emission-line source
relative to the background (Figure 5, showing recent data from the MDM
Observatory 2.4-meter/CCD Spectrograph).
Since the spectral reflectance function of the lunar surface is largely
homogenized by impact mixing of the regolith, more than 99\% of the light in
such a spectrum can be simply ``subtracted away'' by imposing this average
spectrum and looking for deviations from it (Figure 6).
If a TLP radiates primarily in line emission, this factor along with our
ability to reject photons outside the line profile yields a contrast as high as
10,000 times better than the human eye observing the Moon through a telescope.
This could also be done farther into the infrared, for instance we are
preparing to observe the L-band (2.9-4.3~$\mu$m) using SpeX on the NASA
Infrared Telescope Facility in single-order mode, which can deliver $R \la
2000$.

In general observations of this kind might be useful in the infrared for wider
band emission, which is repeatable based primarily on temperature (versus
ionizing excitation as in Paper II, Appendix 1).
Using the HITRAN database to compute vibrational/rotational states for
different molecules, one can see these starting in the infrared (or smaller
wavenumbers for H$_2$O, NH$_3$, CO and CH$_4$), and extending into the optical
for H$_2$O but at least to K-band for NH$_3$ (and intermediate bands for
CO$_2$, CO and CH$_4$).
At least for these molecules, the band patterns are strong and highly distinct.

To be clear, this latter idea requires having an IR spectrograph available at
several minutes notice to follow up on an alert of a TLP (probably found in
imaging).
On a longer timescale, IR spectroscopy might also be useful for the L-band
hydration test outlined above, especially on some of the narrower spectral
features near 3.4 $\mu$m that imaging might overlook, even through narrow-band
filters.

The data cube described above can be sliced in any wavelength to construct a
map of lunar features in narrow or broad bands.
Figure 7 shows that specific surface features can be reconstructed in good
detail and fidelity.

\subsection{Imaging from High Orbit}

Given the constraints on imaging from the vicinity of Earth, it is interesting
to consider the limits and potentials of imaging monitors closer to the Moon.
In general, we will not be proposing special-purpose missions in space-based
remote sensing, and indeed will only mention dedicated missions related to
in-situ exploration of areas affected by volatiles, where special-purpose
investment seems unavoidable.
With in-situ cases, we would perform a more extensive study, so will largely
postpone these discussions to later work concentrating on close-range science.
Here we propose experiments and detectors which might ride on other platforms,
either preceding or in concert with human exploration, and which will
accommodate the same orbits and other mission parameters which might be chosen
for other purposes.
Some of these purposes are not designated priorities for planned missions, but
might prove useful and probably should be considered in the future.
In some cases, we will give rough estimates of project costs based on our prior
experience with similar spacecraft.
These are for discussion only and would need to be re-estimated in detail to be
taken with greater credibility.

An instance of such joint use: does exploration of the Moon imply establishment
of a communications network with line-of-sight visibility from essentially all
points on the lunar surface (excepting those within deep craters, etc.)?
If so, these platforms might also serve as suitable locations for comprehensive
imaging monitoring.
A minimal example of such a network might have a tetrahedral configuration
(with each point typically 60000~km above the surface) with a single platform
at Earth-Moon Lagrange point L1, covering most of the nearside Moon, and three
points in wide halo orbits around L2, each covering their respective portion of
the far side plus a portion of the limb as seen from Earth.
No single satellite will be capable of covering the entire far side, especially
if operation of farside radio telescopes there require a policy of solely
high-frequency communications e.g., via optical lasers.
A single L2 satellite will cover at most 97\% of the far side (subtending
$176^\circ .8$, selenocentrically);
full coverage (not to mention some communications system redundancy) will
require three satellites, plus some means of covering the near side.
With this configuration, the farthest points from each satellite will be
typically $71^\circ$ (in selenocentric angle), hence forshortened due to
proximity to the limb by $\sim 3$ times.
Extensive discussion is underway of using a facility at L1 to aid in transfer
orbits throughout the solar system (Lo 2004, Ross 2006); in that case we should
also consider placing an imaging monitor at L1.

An imaging monitor to improve significantly on Earth-vicinity
capabilities might need to be an ambitious undertaking.
For instance, to acheive 100m FWHM resolution at the sub-satellite point on the
face of the Moon requires an imager of about 4~Gpixels, an aperture $\ga 0.5$~m,
and a field-of-view of 3$^\circ .3$.
Each such monitor, separate from power, downlink, attitude control and other
infrastructure requirements will cost perhaps \$100M.
A stand-alone facility might cost several times more, at each of the several
stations.
Perhaps the system could be cut to a single farside monitor, in a narrow halo
orbit extending beyond the Moon's Earth-shadow, plus some nearside monitoring,
which together could still cover perhaps 95\% of the lunar surface, albeit with
some extreme limb foreshortening.
We also need to ask ourselves at some point if the essential research and
resource exploitation might be confined to the near side.
This is an expensive undertaking, and one that must probably be combined with
other reasons to establish platforms near L1 and L2.
In the meantime, we should accomplish what is possible from the ground. 

If the goal is to discover the source of volatiles for the sake of further
scientific exploration or resource exploitation, however, an investment in
remote sensing, in terms of spatial resolution (or spectral resolution to
discover the substances involved, or temporal resolution to define the behavoir
of the source) makes in-situ reconnaissance and exploration much less
problematic.
A human mission, or a sophisticated robotic mission, could conceivably cost
\$1B, and remote sensing could inform this effort as to where to look in
detail, when dangerous eruptions might occur, and what is the material goal.
Without such information, these investigation is likely to be more
time-consuming, problematic, and perhaps more hazardous.
We concentrate further on remote sensing, even if the proposed expense might be
significant.

\subsection {Surface and Subsurface Radar}

As explained in Paper II, an expectation of water vapor seepage from the lunar
interior should be an ice layer within the regolith about 15m below the lunar
surface.
A remote means of studying this feature would be ground-penetrating radar,
either from the ground or spacecraft platforms.

One should realize that there is significant heritage and as well as plans
involving lunar radar.
The Lunar Sounder Experiment (LSE) on {\it Apollo 17} (Brown 1972, Porcello
1974) operated in both a high-frequency and penetrating radar mode (5, 16 and
260 MHz).
Also planned are the Lunar Radar Sounder (LRE) aboard SELENE (Ono \& Oya 2000:
at 5 MHz (with an option at 1 MHz and 15 MHz), and Mini-RF on the Lunar
Reconnaissance Orbiter, operating at 3 GHz and $\sim$10 GHz.
Finally, of note for comparison's sake in the martian case is MARSIS (''Mars
Advanced Radar for Subsurface and Ionosphere Sounding'' at 1.8, 3.0, 4.0, and
5.0 MHz: Porcello et al.\ 2005).

At 5 MHz ($\lambda = 60$m) the depth of penetration is many kilometers below
the lunar surface, but the spatial resolution is necessarily coarse.
To study the regolith and shallow bedrock, we should choose a frequency closer
to 100-300 MHz.
The Apollo LSE operated for only a few orbits and only close to the equator.
The SELENE LRE runs at lower frequency.
A higher frequency mode is desirable.

The ground-based alternative is useful; lunar radar maps have been made at
40 MHz, 430 MHz, and 8 GHz (Thompson \& Campbell 2005), also 2.3 GHz (Stacy
1993, Campbell et al.\ 2006a, b).
At 8 GHz we are only studying structure of several centimeters within a meter
of the surface.
For 430 MHz we see perhaps $\sim 10$~m inside, and at 40 MHz, 100~m towards the
interior (with attenuation lengths of roughly 10-30 wavelengths).
In practice, better angular resolution at higher frequencies is possible e.g.,
20~m (Campbell et al.\ 2006a, b).
Of course from Earth only the nearside is accessible, and larger angles of
incidence e.g., $\sim 60^\circ$, imply echoes dominated by diffuse scattering
in a way which cannot be modulated.

Use of circular polarization return measurements can be used to test for water
ice (Nozette 1996, 2001) but have been questioned (Simpson 1998, Campbell et
al.\ 2006).
We will not review this debate here, but application of the idea to subsurface
ice is problematic.
It is unclear that this could be accomplished at frequencies of hundreds of MHz
required to penetrate to depths of $\sim$15m, and the more standard technique
(at 13 cm) only performs to depths $\la$1m, where ice sublimation and diffusion
rates are almost certainly prohibitive of accumulation.

Finding subsurface ice has its challenges.
For instance, the dielectric constant $K \approx 3$ for both regolith and water
ice (which is slightly higher), as it is for many relevant mineral powders of
comparable specific gravity e.g., anorthosite and various basalts.
Ice and these substances have similar attenuation lengths, as well.
On the strength of net radar return signal alone, it will be difficult to
distinguish ice from any usual regolith by their mineral properties.
However, in terrestrial situations massive ice bodies reflect little internally
e.g., Moorman, Robinson \& Burgess (2003).
One might expect ice-bearing regions to be relatively dark in radar images,
if lunar ice-infused volumes homogenize or ``anneal'' in this way, either by
forming a uniform slab or by binding together regolith into a single,
uniform $K$ bulk.

On the other hand, hydrated regolith samples have $K$ values much higher than
unhydrated ones (by up to an order of magnitude), as well as attenuation
lengths even more than an order of magnitude shorter (Chung 1972).
This hydration effect is largest at lower frequencies, even below 100 MHz.
One might suspect that significant water ice might perturb the chemistry of the
regolith significantly, which might even increase charge mobility as in a
solution, which appears to invariably drive up $K$, and conductivity even more,
increasing the loss tangent: conductivity divided by $K$ (and the frequency).
One should expect a reflection passing into this high-$K$ zone, but this
depends strongly on the details of the suddenness of the transition interface.

Of particular interest is the radar map at 430 MHz (Ghent et al.\ 2004) of the
Aristarchus region, site of roughly 50\% of TLP and radon reports.
The 43-km diameter crater is surrounded by a low radar-reflectivity zone some
150~km across, particularly in directions downhill from the Aristarchus plateau
onto Oceanus Procellarum.
In general the whole plateau is relatively dark in radar, occasionally
interrupted by bright crater pock-marks and Vallis Schr\"oteri.
In contrast the dark radar halo centered on Aristarchus itself is uniquely
smooth, indicating that it was probably formed or modified by the impact
itself, a few hundred million years ago.
This darkness might be interpreted as higher loss tangent, consistent with the
discussion in the previous paragraphs, or simply fewer scatterers (Ghent et
al.~2004) i.e., rocks of approximately meter size; it is undemonstrated why the
latter would be true in the ejecta blanket of a massive impact especially given
the bright radar halo within 70 km of the Aristarchus center.
Ghent et al.~(2005) show that other craters, some comparable in size to
Aristarchus, have dark radar haloes, but none so extended.
The region around Aristarchus has characteristics that might be expected from
subsurface ice redistributed by impact melt: dark, smooth radar-return,
spreading downhill but otherwise centered on the impact; this should be
expected to be confused, at least, with the dark halo effect seen around some
other impacts.
It seems well-motivated to search for similar dark radar areas around other
likely outgassing sites, particularly ones not associated with recent impacts;
unfortunately, the foremost candidate for such a signature is competing with
such an impact, Aristarchus, which can be expected to produce its own confusing
effect.

We would propose that radar at frequencies near hundreds of MHz be considered
for future missions, in a search for subsurface ice.
This is a complex possibility that we will not detail here, that must be
weighed against the potential of future ground-based programs.
In particular, the near side has been mapped at about 1~km resolution for 70~cm
wavelength (Campbell et al.~2007), this could be improved with an even more
intensive ground-based program, or from lunar orbit.
Orbital missions can be configured to combine with higher frequencies and
different reception schemes to provide better spatial resolution, deal with
ground clutter, and varying viewing angles.
A lunar orbiter radar map would be less susceptible to interference speckle
noise, which will likely require long series of pointings to be reduced from
the ground.
In combination with an optical monitor, a GHz-frequency radar might produce
detailed maps in which changes due to TLPs might be sought, and might be then
correlated with few-hundred MHz maps to aid in interpretation in terms of
volatiles.

At shorter wavelengths one should consider mapping possible changes in surface
features due to explosive outgassing, which Paper II hints might occur
frequently on scales excavated over tens of meters, and expelled over hundreds
or thousands of meters.
Again, earth-based observations suffer from speckle, but planned observations
by the {\it Lunar Reconnaissance Orbiter (LRO)} Mini-RF (Mini Radio-Frequency
Technology Demonstration - Chin et al.\ 2007) at 4 and 13~cm might easily make
valuable observations of this kind.
Both modes scan in a swath $\sim$5 km wide, which would make comprehensive
mapping difficult, but would mesh well with the event resolution from a
ground-based optical monitor.
A ``before'' and ``after'' radar sequence meshed with an optical monitoring
program would likely be instructive as to how outgassing and optical transients
actually interact with the regolith.

%
%

\subsection {Monitoring from Low Lunar Orbit}

\subsubsection {Planned Optical Imaging}

Several upcoming missions will carry high-resolution optical imagers, each of
which will be capable of mapping nearly the entire lunar surface e.g.,
{\it Chang'e-1} CCD imager (Yue et al.\ 2007),
{\it SELENE} Spectrometer/Multiband Imager (LISM/MI) (Ohtake et al.\ 2007),
{\it LRO} Camera (LROC) (Robinson et al.\ 2005),
and
{\it Chandrayaan-1} Moon Mineralogy Mapper (MMM) (Pieters et al.\ 2006),
typically at tens to hundreds of meters resolution.
In particular the MI/SP will usefully observe at 20m resolution the
pyroxene near-IR band that can indicate the exposure of fresh surface,
as can the MMM (albeit at 280m resolution).
All of these are sensitive at blue wavelengths which can also indicate surface
age.
The LROC and MMM will repeatedly map each point on the Moon, not in any way
sufficient to be considered realtime monitoring of transients, but sufficient
to allow frequent sampling on timescales of a lunation.
This allows an interesting synergy with ground-based monitors since they can
highlight sites of activity for special analysis.
Furthermore, LROC has a high resolution pointed mode which might provide
sub-meter information in areas where TLPs have been recently detected, hence
excellent sampling on the scales that we suspect will be permanently effected,
perhaps in a ``before'' and ``after'' sequence.
At any given time, any these four spacecraft have a roughly 10\% chance of at
least one of them being in view of a particular site above its horizon; it
would be fascinating (but perhaps too logistically difficult) if a program
could be implemented wherein spacecraft could be alerted to image at high
resolution a TLP site in real time during an event.  

%
%
%

\subsubsection {Alpha-Particle Spectrometry}

In order to study outgassing directly, we need instruments at or near the lunar
surface.
In the case of $^{222}$Rn, the thermal velocity is typically $v \approx 150$~m
s$^{-1}$, so typical ballistic free flight occurs over $d = v^2 / g = 7$~km.
Over its half-life of 3.8~d, a $^{222}$Rn atom travels typically 50000~km in a
random walk that wanders from the source only a few hundred km before decaying
(or sticking to a cold surface).
Thus the alpha particles must be detected in much less than a day after
outgassing, or the $^{222}$Rn signal disperses by an amount that makes
superfluous placing the detector less than a few hundred km above the lunar
surface, except for $r^{-2}$ sensitivity considerations.

Three alpha-particle spectrometers have observed the surface of the Moon, but
for relatively brief periods of time.
The latitude coverage was severely limited on {\it Apollo 15} ($\mid Lat \mid 
\la 26^\circ$ for 145 hours) and {\it Apollo 16} ($\mid Lat\mid \la 5^\circ$,
128~h).
{\it Lunar Prospector's} Alpha Particle Spectrometer covered the entire Moon,
over 229 days spanning 16 months, but was partially damaged (one of five
detectors) upon launch and suffered a sensitivity drop due solar activity
(Binder 1998).
{\it Apollo 15} observed two outgassing events (from Aristarchus and Grimaldi),
{\it Apollo 16} none, and {\it Lunar Prospector} two sources (Aristarchus and
Kepler), although the signals from these last sources were integrated over the
mission duration.
In addition, Apollo and {\it Lunar Prospector} instruments detected an
enhancement at mare/highlands boundaries from daughter product $^{210}$Po,
indicating $^{222}$Rn leakage over approximately the previous century.

The expected detection rate for a single alpha-particle spectrometer in a polar
orbit and without instantaneous sensitivity problems, might be grossly
estimated from these data.
The {\it Apollo 16} instrument covered a sufficiently small fraction ($\sim
12$\%) of the lunar surface so that we will not consider it, whereas {\it
Apollo 15} covered about 37\%.
These missions were in orbit $\sim$6~d apiece, and considering the $^{222}$Rn
lifetime thereby were sensitive to events (at $>$10\% full sensitivity) for
$\sim 18$~d.
{\it Lunar Prospector} covered the entire lunar surface every 14~d, hence
caught events typically at 28\% instantaneous full strength (minimum 8\%),
however, by averaging over the mission diluted this by an factor $\sim$20-30.
These data are consistent with a picture in which Aristarchus produces an
outgassing event 1-2 times per month at the level detectable by {\it Apollo
15}, and by {\it Lunar Prospector} when integrated over the mission.
Apparently other sites such as Grimaldi and Kepler collectively are about
equally active as Aristarchus, together all sites might produce 2-4 events per
month at the sensitivity level of {\it Apollo 15}.
This level of activity is consistent with the statistics of TLPs constrained in
Paper I.

A new orbiting alpha-particle spectrometer with a lifetime of a year or more
and an instantaneous sensitivity equal to that of {\it Apollo 15}'s detector
would likely produce a relatively detailed map of where outgassing occurs on
the lunar surface, separate from any optical manifestation.
This is likely an important test for many of the procedures mentioned above,
which are critically dependent on the outgassing/optical correlation.
This must be examined in further detail, because there are many ways in which
one might imagine that gas issues from the interior, thereby producing radon,
without a visible manifestation, either due on one extreme to such rapid
outgassing that previous events have cleared the area of regolith that might
interact with gas on its way to the vacuum, or due to seepage sufficiently slow
to trap water (and perhaps other gasses by reaction) in the regolith, and too
slow to perturb dust at the surface.
Radon, an inert gas that will not freeze or react on its way to the surface, is
more likely to escape the regolith to be detected, regardless.

The Alpha-ray Detector (ARD) onboard {\it SELENE} (Nishimura et al.\ 2006)
promises to be $\sim$ 25 times more sensitive than the Apollo Alpha Particle
Spectrometers, with a mission lifetime of one year or more, in a polar orbit.
This, in conjunction with an aggressive optical monitoring program (as in
Section 2.1), holds the prospect of extending the TLP/$^{222}$Rn-outgassing
correlation test from Paper I to a dataset of order 10 times larger.
This would likely serve as a significant advance in understanding their
connection, but it is probably best to consider what a following generation
alpha-particle spectrometer study might entail.

To insure better sensitivity coverage two such detectors in complementary
orbits would cover the lunar surface every 1.8 half-lives of $^{222}$Rn.
This may nearly double the detected sample.
Unless the alpha-particle detectors are constructed with a veto for solar
wind particles, it is best to avoid active solar intervals.
We will exit the solar minimum probably by year 2008, with the next starting by
about 2016.
On the other hand, some of the lack of sensitivity to lunar alpha particles and
elevated solar particle background count on {\it Lunar Prospector} was due in
part to it being spin-stabilized.
If detectors on a future mission were kept oriented towards the lunar surface
and shielded from solar wind to the extent possible, the Apollo results
indicate that prompt $^{222}$Rn outburst detection at good sensitivity is
possible.
Beyond this, extending the mission(s), of course, will help, and the best
approach might be to develop a small alpha-spectrometer package that might
easily fly on any extended low-orbital mission.

\subsubsection {On-Orbit Mass Spectrometry}

The radioactive decay delay in alpha-particle detection insures that a
reasonable number of orbiting detectors can have near unit efficiency.
This is not the case for prompt detection of outgassing e.g., by mass
spectrometers.
An instantaneous outburst seen 100~km away will undergo a dispersion of only a
few tens of seconds in arrival time.
The detectors must either be very sensitive or densely spaced, and prepared to
measure and analyze what they can in these short time intervals.
This is a problem for Apollo-era instruments e.g., the {\it Apollo 15} Orbital
Mass Spectrometer Experiment (OMSE - Hoffman \& Hodges 1972) required 62s to
scan through a factor of 2.3 in mass (12 to 28, or 28 to 66~AMU).

Total amount of outgassing is in the range of many tons per year, and with
perhaps tens of outbursts per year, the mass fluence of particles from a single
outburst seen at a distance of 1000~km is approaching $10^{12}$~cm$^{-2}$ AMU.
While a burst on the opposite side of the Moon will not be detected and/or
properly interpreted, one that can be seen by a few detectors would be very
well constrained.

The specific operational strategies of these detectors is paramount.
For example consider an event at 1000~km distance, which will spread over $\sim
500$s in event duration.
A simple gas pressure gauge will not be overwhelmingly sensitive, in that even
with an ambient atmosphere that is not unusual e.g., number density $n \approx
10^4-10^5$ cm$^{-3}$ (varying day/night e.g., Hodges, Hoffman \& Johnson 2000),
the background rate of collisions over 500~s amounts to an order of magnitude
or more than the particle fluence than for a typical outgassing outburst,
assuming $\sim 20$ AMU particles in the outburst.
Since interplanetary solar proton densities can change by amount of order unity
in an hour or less (e.g., McGuire 2006), pressure alone is not likely to be a
useful event tracer.

A true mass spectrometer is useful in part by subdividing the incoming flux, in
mass, obviously, but also in direction, thus decreasing the effective
background rate.
The disadvantage of this approach in the past has been that it cannot cover the
entire parameter range of this subdivision at once, so must scan in atomic mass
or direction, or must always accept a significantly limited range.
For a short burst, this means that mass components may not be examined during
the event, or that events might be missed due to detectors pointing in the
wrong direction.
For Apollo-era detectors, these problems, particularly the former, were
significant.
We would prefer to operate a mass spectrometer operating continuously over a
significant mass range, with ballistic trajectory reconstruction over a large
incoming acceptance solid angle.
We will return to this concept below.

First, let us discuss low-orbit platforms.
We will not propose special purpose probes of the atmosphere
alone, but there are other reasons for dense constellations of lunar
satellites, most prominently a lunar global positioning system (GPS).
Terrestrial systems in operation (GPS) and planned (Galileo, Beidou and
GLONASS: GLObal NAvigation Satellite System) are typically 25-30 satellites at
orbital radii $\sim$25000~km.
Around the Moon this could be much lower, $\sim 8000$~km, and with fewer
satellites, $\sim$12, which would put satellites within $\sim$7000~km of a
surface outburst.
This is compared to $\sim 100$~km for Apollo.
Scaling the sensitivity of the {\it Apollo 15} OMSE (Hodges et al.~1973), a
detector on a GPS would be sensitive (at the 5$\sigma$ level) to an
instantaneous outburst of about 50000~kg (and more depending on the details of
non-$r^{-2}$ propagation effects).
This is insufficient sensitivity to detect outgassing events.
One needs a lower orbit (or much more sensitive detectors, by three orders
of magnitude).

It is unclear if a lower-orbit GPS system, while more favorable for an add-on
mass spectrometer array, would serve its navagational purpose.
A GPS/mass spectrometer constellation only 1000~km above the lunar surface
could likely be made sufficiently sensitive for gas outburst monitoring, nearly
continuously.
Such a low orbit makes GPS more difficult, require several more
satellites, and increasing the effects of mascons on their orbit.
This requires further modelling.

Nonetheless, we should consider other science instrumentation on a lunar GPS.
High-resolution imaging from $\sim 8000$~km radius could be 10$\times$ finer
($\sim 10~$m) than platforms at L1 or near L2.
Covering the Moon at this resolution would require $\sim 10^{12}$ pixels, which
might allow mapping occasionally, but only crude monitoring temporally.
Still, if one-third of lunar GPS platforms were equipped with a prompt,
high-resolution imager, any portion of the lunar surface could be imaged during
the course of a surface event.
If an event is observed from the ground or from L1/L2, it could be detailed at
10~m or even higher resolution.
This imager network should establish an atlas of global maps (at various
illumination conditions) to serve as a ``before'' image in this comparison (as
well as allowing a wealth of other studies).
By allowing transient events to be studied at $\la 10$~m resolution, this sets
the stage for activity to be isolated at a sufficiently fine scale for in-situ
investigations that would thereby be targetted and efficient in localization.

Returning to mass spectrometry, it is clear that there are two separate modes
for gas propagation above the lunar surface, neutral and ionized, and that a
significant amounts are seen in both (Vondrak, Freeman \& Lindeman 1974, Hodges
et al.~1972), at a rate of one to hundreds of tonne~y$^{-1}$ for each process.
There is some possibility that a large portion of the ionized fraction might be
molecular in nature (Vondrak et al.~1974).

For neutral atoms more massive than H or He, their thermal escape lifetime is
sufficiently long that they have ample time to migrate across the lunar surface
until they stick in a shadowed cold-trap.
Furthermore, the ionized component will predominently follow the electric field
embedded in the solar wind, which tends to be oriented perpendicular to the
Sun-Moon vector and hence frequently pointing from the sunrise terminator into
space.
For these two reasons the best location to monitor outgassing is a point above
the sunrise terminator, presumably on a low-orbit platform.
Note that there is some degeneracy between the timing information recorded by
a particle detector on such a satellite between the episodic behavior of
particle outgassing versus the motion of the spacecraft at $\sim 1.6$ km
s$^{-1}$.
The ideal situation would be to triangulate such signals with more than one
platform.
Such an experiment is not trivial, but there are alternatives, explored below.

For a low lunar orbit to be ``low maintenance'' i.e., require few
corrections due to mascon perturbations, it should be at one of several special
``frozen orbit'' inclination angles $i = 27^\circ$, $50^\circ$, $76^\circ$ or
$86^\circ$ (e.g., Ramanan \& Adimurthy 2005).
However, we want to maintain a position over the terminator, using a
sun-synchronous orbit, which requires a precession rate $\omega_p = 0.99^\circ
d^{-1} = 2 \times 10^{-7}$ rad s$^{-1}$.
Natural precession due to lunar oblateness is determined by the gravitational
coefficient $J_2 = (2.034 \pm 0.001) \times 10^{-4}$ (Konopliv et al.~1998)
according to $\omega_p =-(3 a^2 J_2 \omega cos ~i)/(2 r^2)=-(3 a^2 J_2 sqrt{GM}
cos~i)/(2 r^{7/2})$, where $a$ is the lunar radius,
$\omega$ the orbital angular speed, $M$ the lunar mass and $r$ the orbital
radius.
(The precession caused by Earth is 1000 times smaller, and 60000 times smaller
for the Sun.)~ 
One cannot effectively institute both conditions, however, since the maximum
inclination orbit with $\omega_p = 2 \times 10^{-7}$ s$^{-1}$ occurs at
$47^\circ$ (or else the orbit is below the surface).
While an orbit at $i = 27^\circ$ is stable (at $r = 1876$~km, 138~km above the
surface) and has the correct precession rate, it spends most of its time away
from the terminator.

In contrast, at $i = 87^\circ$, $\omega_p = 1.5 \times 10^{-8}$ s$^{-1}$, and
the spacecraft needs to accelerate continuously only $a = 0.3$~mm s$^{-2}$ to
place it into sun-synchronous precession.
This is nearly the same as the thrust provided by the Hall-effect ion engine on
{\it SMART-1}
(and corresponds to an area per mass of 330 cm$^2$ g$^{-1}$ under the influence
of solar radiation pressure.)~
While it is not apparent that an ion engine would be the best choice for a
platform with mass and ion spectrometers, this illustrates the small amount of
impulse need to maintain this favorable orbit, comparable to station-keeping
in many non-frozen orbits.
In truth, the most efficient location to apply this acceleration is only near
the poles, so a slightly more powerful thruster might be needed.
Since, time-averaged, this perturbed orbit still lands in a frozen-orbit zone,
it should still be relatively stable in terms of radius.
We would propose that a instrumented platform in this driven, sun-synchronous
polar orbit would be ideal for studying outgassing signals near the
terminators.
%
%

There is an interesting synergy between this outgassing monitor platform and
another useful investigation from a similar satellite(s), although not
necessarily simultaneously.
An outstanding problem is gravitational potential structure of the Moon,
particularly the far side (where satellite orbits cannot be monitored from
Earth).
With the inclusion of the the 562-day {\it Lunar Prospector} data set (Konopliv
et al.\ 2001) the error is typically 80 milligals on the far side (corresponding
to surface height errors of about 25 m) versus 10 milligals in the near-side
potential.
Also the limiting harmonic is of order 110 approximately on the near side, and
only order 60 on the far side ($\approx$ 200 km resolution).

In contrast, the GRACE (Gravity Recovery and Climate Experiment) can define the
geodesy of Earth at much better field and spatial resolution, a few milligals
at about order 200 (Tapley et al.\ 2005 - one year of data), using a double
satellite at $\sim$500 km above the Earth in polar orbit, with the separation
($\sim$200 km) between the two components carefully monitored (by laser
interferometer for the proposed GRACE follow-on mission - Watkins et al.\ 2006,
or in the microwave K-band for GRACE itself).
Such a satellite pair in lunar orbit would improve our knowledge of the farside
field by orders of magnitude, determined independent of Earth-based tracking
measurements, and in general make the accuracy and detail of lunar potential
mapping much closer in quality to mineralogical mapping already in hand.
One interesting question this might address is whether mascons extend to much
smaller scales than currently known.
While this mapping is underway, one could use outgassing monitors on board to
look for outbursts, and when the geodetic mission is complete, drive the
satellites into a polar, sun-synchronous orbit above the terminator.
Depending on the type of monitors imployed, forcing sun-synchronous precession
by chemical, ion or even solar-sail propulsion may or may not interfere;
neutral-gas spectrometers may be compatible with ion drives while charged
species trajectories might be perturbed, for instance.

Maintaining $a = 0.3$~mm~s$^{-1}$ for a 100~kg spacecraft requires 20 kg
month$^{-1}$ of chemical propellant (exhaust velocity of 4000 m s$^{-1}$)
versus 2.5 kg month$^{-1}$ of ion propellant (30000 m s$^{-1}$).
For a 100 kg spacecraft a solar sail about 30m in radius would be required.
None of these solutions are so easy that they do not inspire a search for
alternatives, and their non-gravitational acceleration would mean that they
could take place only after (or before) any geodesic mission phase.
Furthermore, ion propulsion and probably chemical propulsion would tend to
interfere with mass spectrometry.
These should be traded against other possibilites e.g., several small probes on
various orbital planes at $i = 87^\circ$, rather than one or two
sun-synchronous platforms.
%

The fact that there would be an outgassing detectors on each platform would
make temporal/spatial location of specific outbursts more unambiguous, aided by
differences in timing and signal strength at the two moving platforms, at least
for neutral species.
The timing difference will give an indication of the distance difference to the
sources, with the source confined to the hyperboloid $x^2/a^2-y^2/(e^2-a^2)=1$
where $x$ is the distance along the line connecting the two satellites, with
the origin at the half-way point between them, and $y$ is the distance
perpendicular to this line.
The distance between the two satellites is given by $2e$ and the difference in
distance between the source and the first satellite versus the source and the
second is $2a$.
There is still a left/right ambiguity in event location to be resolved by
detector directionality, and better directional sensitivity would add a helpful
overconstraint on the measurement.

\newpage
\section {In-Situ and Near Surface Exploration}

Our research group\footnote{AEOLUS: ``Atmosphere seen from Earth, Orbit and the
LUnar Surface'' - see Crotts et al.\ 2007} is developing ways to efficiently
transfer the insight gained from a program of remote sensing to a program of
in-situ research involving the lunar surface.
I would like to emphasize a few key points already becoming apparent.

The neutral fraction from lunar outgassing need not respect the correlation
with lunar sunrise; a detector giving enough prompt information about
outgassing might be invaluable.
Neutral gas emitted on the day side is free to bounce ballistically until
either sticking to a cold surface or escaping (either due to ionization or by
reaching the high-velocity maxwellian tail).
A highly desirable monitor of this activity would be a mass spectrometer
capable of simultaneously accepting particles in a wide range of masses
e.g., $\sim 10-100$ a.m.u., and reconstructing incoming particle trajectories
and velocities to allow the locus of outgassing to be reconstructed (at least
within hundreds or thousands of km).

In addition to tracking the sunrise terminator outgassing signal, such a mass
spectrometer would be able to monitor wide areas of the Moon for prompt neutral
outburst signals from point sources, and therefore the instrument should be
placed in the vicinity of known outgassing sites to establish which species
succeed in propagating to the regolith surface.
The suggested ground-based approaches provides this rough localization,
buttressed by the low-orbital outgassing detectors.

At some point the identification of a good tracer gas to act as a proxy for
endogenous emission would be highly valuable in simplification of outgassing
alert monitors not required to scan entire mass ranges.
Now it is unclear what that gas should be.
It is true that $^{222}$Rn seems to be highly correlated with optical
transients, but the relationship between radiogenic gas emission and that of
volcanic emission is uncertain.
Besides, while usefully radioactive, radon is a very minor constituent.
Radiogenic $^{40}$Ar is more abundant, and episodic, but its relation to
volcanic gas is uncertain (as is its correlation to optical transients).
The most reliable observed molecular atmospheric component is CH$_4$, but it is
likely to derive in large part from cometary/meteoritic impacts and is somewhat
unnatural to expect from the oxygen-rich interior.
Water suffers from the situation described in Paper II in which a large
fraction of any large, endogenous source might never propagate gas to the
surface, making it an unreliable tracer.
Even while endogenous water of nearly certain volcanic origin has been found in
glasses likely derived for the deep interior (Saal et al.\ 2007), CO$_2$ is
absent.
The limits on CO are more unclear, as are those for oxides of nitrogen.
The first mass spectrometer probes should be designed to clarify this
situation.

To place these monitors on the surface, one may exploit human exploration
sorties, which will be relatively infrequent and potentially concentrated in
sites of just a very few bases.
I reiterate that another concern is the contamination that each of the missions
will produce, concentrated primarily near the landing site itself.
It is evident that by the deployment of LACE on the final Apollo landing that
the outgassing environment was contaminated by a large contribution of 
anthropogenic gas, and that these vehicles in a new epoch of human exploration
will deliver many tens of tons per mission of gases to the lunar surface of
composition relevent to species suspected from a potential endogenous volcanic
component, a level of contamination comparable to the potential annual output
of such gases from endogenous sources.

The Constellation spacecraft consist of Orion, carrying about 10 tonne of
N$_2$O$_4$ (nitrogen teroxide) and CH$_3$N$_2$H$_3$ (monomethyl hydrazine)
propellant, and LSAM, propelled by liquid oxygen and nitrogen.
The Orion fuel mix produces N$_2$, CO$_2$ and H$_2$O and the LSAM exhausts
water.
Depending on the orientations and trajectories of the spacecraft when thrusting
they will deposit about 20 tonnes of mostly water to the surface, where most
will remain for days (up to about one lunation).
During the course of the Return to The Moon, measurements of at least these
three product molecules will be suspect, since in fact their signal will
disappear completely over successive lunations.

In many respects the surface layer of regolith should be considered as a
planet-sized sorption pump coupling the atmosphere, across which gases are
free to propagate (and exit the system if they are ionized or low-mass), and
the lower regolith, which is cold ($\sim 250$K) and relatively impermeable.
Gas in the atmosphere can be delivered to the surface where, if it penetrates a
few cm, enters a region in which particle mobility slows considerably and where
it essentially becomes entrained in the time-averaged signal of endogenous gas
(radiogenic or volcanic) that is leaking from greater depth.
(Indeed, since the temperature increases inwards, gas reaching this colder zone
preferential migrates to greater depths.)~
Furthermore, once gas from the interior reaches the outer few cm of regolith
subject to large temperature swings, it is likely to escape into the vacuum.

There is a scientific premium, therefore, to delivering surface monitors to
their site without delivery of many tons of anthropogenic gas, annd for this
purpose one might consider small, parasitic landing rockets that deliver an
experiment package from the Orion or LSAM human exploration vehicles to the
vicinity of the surface, but transition to a low-contamination soft lander
system such as an airbag.
This is an established, low-cost technology with extensive heritage (from the
Ranger Block 2 lunar probes to the highly successful Mars Exploration Rovers)
and might easily be the landing technique of choice for small lunar surface
packages.
On small ($\la$ tens of km) scales, robotic rovers are less prone to sowing
contamination when delivering detector packages across the surface.

When human exploration turns towards study of lunar outgassing sites the
primary challenge may be converting the lower spatial resolution information
obtained at Earth or lunar orbit into meter or 10-meter scale intelligence
regarding where to initiate in-situ exploration.
The transitional technologies to bridge this gap consist of local networks of
sensors that map area on the scale of a 1 km or 100 m to resolutions of 1-10 m
using various techniques: local ground-penetrating radar, local seismic arrays,
directional and ground-sniffing mass-spectrometers that work to localize, and
another technique we propose to investigate: intensive laser grids that
densely populate the space above the patch of surface in question with
lines of sight sampling strong transitions of some predominant species e.g.,
an infrared vibrational/rotational transition if molecules are discovered in
quantity.

The details of the ideas promulgated in this section are beyond the scope of
the current paper and will be presented in a larger document currently in
preparation.

If the reader will allow a personal statement, I am not easily swayed into
writing research papers based on data of the uncertain quality of those seen
in Papers I and II, but this is the nature of the field.
It has been the purpose of this investigation not only to clarify the
implications of existing data, which I think it has done, but also to
understand the range of interesting possibilities of phenomena consistent with
these data and ask how we should proceed to investigate them, cognizant that
many of our actions have implications in terms of disturbing the environment
that we care to assay.
We need to access which interesting questions need to be addressed, given the
state of our ignorance, and consider how to proceed.
I hope and intend that these works have advanced the discussion significantly.

\section{Discussion and Conclusions}

The phenomena that we have been studying are subtle, and many important
aspects may be highly covert.
The above-surface signals of outgassing of radiogenic endogenous sources is
fairly clear, but gas of more magmatic origin, while possibly present, needs
further study to be absolutely confirmed.
Activity associated with Apollo landings easily dominated with anthropogenic
gas production the activity in molecular species that might trace residual
lunar magnatism.
Apollo-era and later data were insufficiently sensitive to establish the level
of outgassing beyond $^{222}$Rn, $^{20}$Ne and isotopes of Ar, plus He,
presumably, but did detect molecular gas, particularly CH$_4$, but of
uncertain origin.
It is important to assess how we can advance the Apollo-era understanding.
Consistent with these molecular gas outflows, and perhaps traced by optical
transients, there is a range of possible phenomena that have interesting
possible scientific consequences and might easily be useful in terms of
resource exploitation for human exploration.
While this amount of volatile production is inconsequential on the scale of
the geology of the Moon as a whole, and is poorly constrained by any
measurement of current or previous volatiles, even in returned surface
samples, it is still capable of massively altering the environment locally in
ways which should be investigated in a timely way.
We could learn a great deal from the current production of volatiles and their
accumulation over geologic timescales in an extraterrestrial environment so
easily explored.

The salient facts from the above treatment is that for many years yet 
monitoring for optical transients will still be best done from the Earth's
surface, even considering the important contributions that will be made by
lunar spacecraft probes in the next several years.
These spacecraft will be very useful in evaluating the nature of transient
events in synergy with ground-based monitoring, however.
Given the likely behavior of outgassing events, it is unclear that in-situ
efforts alone will necessarily isolate their sources within significant
winnowing of the field by remote sensing.
Early placement of capable mass spactrometers of the lunar surface, however,
might prove very useful in refining our knowledge of outgassing composition,
in particular a dominant component that could be used as a tracer to monitor
outgassing activity with more simple detectors.
This must take place before significant pollution by large spacecraft, which
will produce many candidate tracer gasses in their exhaust.

We do not know enough now to discuss the potential implications of this line
of research in terms of resources for human exploration, or even in terms of
prebiologic chemistry on the Moon and for tenuous endogenous outgassing and
atmospheric interactions with the regolith on other bodies, but all of these
are interesting, new avenues of such research.
It is crucial that exploration of these issues progress while we have a
pristine lunar surface as our laboratory.

\section{Acknowledgements}

I would much like to thank Alan Binder and James Applegate, as well as Daniel
Savin, Daniel Austin and the other members of AEOLUS (``Atmosphere as seen from
Earth, Orbit and LUnar Orbit'') for helpful discussion.

\newpage

\noindent
{\bf References:}

\noindent
Adams, J.B.\ 1974, JGR, 79, 4829.

\noindent
Akhmanova, M.V., Dementev, B.V., Markov, M.N.\ \& Sushchinskii, M.M.\ 1972,
Cosmic Research, 10, 381

\noindent
Binder, A.B.\ 1998, Science, 281, 1475;
also see video interview, http://lunar.arc.nasa.gov/results/alres.htm

\noindent
Brown, W.E., Jr.\ 1972, Earth Moon Plan., 4, 133.

\noindent
Buratti, B.J., McConnochie, T.H., Calkins, S.B., Hillier, J.K.\ \&
Herkenhoff, K.E.\ 2000, Icarus, 146, 98.

\noindent
Campbell, B.A., Campbell, D.B., Margot, J.-L., Ghent, R.R., Nolan, M., Carter,
L.M., Stacy, N.J.S.\ 2007, Eos, 88, 13.

\noindent
Campbell, D.B., Campbell, B.A., Carter, L.M., Margot, J.-L.\ \& Stacy,
N.J.S.\ 2006, Nature, 443, 835.

\noindent
Campbell, B.A., Carter, L.M., Campbell, D.B., Hawke, B.R., Ghent, R.R.\ \&
Margot, J.-L.\ 2006, Lun.\ Plan.\ Sci.\ Conf., 37, 1717.

\noindent
Charette, M.P., Adams, J.B., Soderblom, L.A., Gaffey, M.J.\ \&
McCord, T.B.\ 1976, Lun.\ Sci.\ Conf., 7, 2579.

\noindent
Chin, G., et al.\ 2007, Lun.\ Plan.\ Sci.\ Conf., 38, 1764.

\noindent
Chung, D.H.\ 1972, Earth Moon \& Plan., 4, 356.

\noindent
Crotts, A.P.S.\ 2007, Icarus, submitted (Paper I).

\noindent
Crotts, A.P.S.\ \& Hummels, C.\ 2007, ApJ, submitted (Paper II).

\noindent
Crotts, A.P.S.\ 2007, et al.\ 2007, Lun.\ Plan.\ Sci.\ Conf., 28, 2294.

\noindent
Dollfus, A.\ 2000, Icarus, 146, 430.

\noindent
Dzhapiashvili, V.P.\ \& Ksanfomaliti, L.V.\ 1962, The Moon, IAU Symp.\ 14,
(Academic Press: London)

\noindent
Eliason, E.M., et al.\ 1999, Lun.\ Plan.\ Sci., 30, 1933

\noindent
Farr, T.G., Bates, B., Ralph, R.L.\ \& Adams, J.B.\ 1980, Lun.\ Plan.\ Sci.,
11, 276.

\noindent
Fried, D.L.\ 1978, Opt.\ Soc.\ Am.\ J., 68, 1651.

\noindent
Garvin, J., Robinson, M., Skillman, D., Pieters, C., Hapke, B.\ \& Ulmer,
M.\ 2005, $HST$ Proposal GO 10719.

\noindent
Ghent, R.R., Leverington, D.K., Campbell, B.A., Hawke, B.R.\ \& Campbell,
D.B.\ 2004, Lun.\ Plan.\ Sci.\ Conf., 35, 1679.

\noindent
Ghent, R.R., Leverington, D.K., Campbell, B.A., Hawke, B.R.\ \& Campbell,
D.B.\ 2005, JGR. 110, doi: 10.1029/2004JE002366.

\noindent
Hazen, R.M., Bell, P.M.\ \& Mao, H.K.\ 1978, Lun.\ Plan.\ Sci., 9, 483.

\noindent
Hodges, R.R., Jr., Hoffman, J.H.\ \& Johnson, F.S.\ 1973, Lun.\ Sci.\ Conf., 4,
2855.

\noindent
Hodges, R.R., Jr., Hoffman, J.H.\ \& Johnson, F.S.\ 1974, Icarus, 21, 415.

\noindent
Hodges, R.R., Jr., Hoffman, J.H., Yeh, T.T.J.\ \& Chang, G.K.\ 1972, JGR, 77,
4079.

\noindent
Hoffman, J.H.\ \& Hodges, R.R., Jr.\ 1972, Lun.\ Sci.\ Conf., 3, 2205.

\noindent
Konopliv, A.S., Asmar, S.W., Carranza, E., Sjogren, W.L.\ \& Yuan, D.N.\ 2001,
Icarus, 150, 1

\noindent
Konopliv, A.S., Binder, A.B., Hood, L.L., Kucinskas, A.B., Sjogren, W.L.\ \&
Williams, J.G.\ 1998, Science, 281, 1476

\noindent
Law, N.M., Mackay, C.D.\ \& Baldwin, J.E.\ 2006, A\&A, 446, 739.

\noindent
Lebofsky, L.A., Feierberg, M.A., Tokunaga, A.T., Larson, H.P.\ \& Johnson,
J.R.\ 1981, Icarus, 48, 453

\noindent
Lipsky, Yu.N.\ \& Pospergelis, M.M.\ 1966, Astronomicheskii Tsirkular, 389, 1.

\noindent
Lo, M.W.\ 2004, in ``Proc.\ Internat'l Lunar Conf.\ 2003, ILEWG 5'' (Adv.\ in
Astronaut.\ Sci., Sci.\ \& Tech. Ser., Vol.\ 108), eds.\ S.M.\ Durst et al.
(Univelt: SanDiego), p.\ 214.

\noindent
Markov, M.N., Petrov, V.S., Akhmanova, M.V.\ \& Dementev, B.V.\ 1979, in {\it
Space Research, Proc.\ Open Mtgs.\ Working Groups} (Pergamon: Oxford), p.\ 189.

\noindent
McGuire, R.E.\ 2006 ``Space Physics Data Facility'' - NASA Goddard Space Flight
Center:  http://lewes.gsfc.nasa.gov/cgi-bin/cohoweb/selector1.pl?spacecraft=omni

\noindent
Moorman, B.J., Robinson, S.D.\ \& Burgess, M.M.\ 2003, Permafrost \&
Periglac.\ Proc., 14, 319.

\noindent
Nishimura, J., et al.\ 2006, Adv.\ Space Res., 37, 34.

\noindent
Nozette, S.\ et al.\ 1996, Science, 274, 1495.

\noindent
Nozette, S.\ et al.\ 2001, JGR, 106, 23253.

\noindent
Ono, T.\ \& Oya, H.\ 2000, Earth Plan.\ Space, 52, 629.

\noindent
Picardi, G., et al.\ 2005, Science, 310, 1925.

\noindent
Pieters, C.M., et al.\ 2005, in {\it Space Resources Roundtable VIII},
Lun.\ Plan.\ Inst.\ Contrib., 1287, 73.

\noindent
Pieters, C.M., et al.\ 2005, 
http://moonmineralogymapper.jpl.nasa.gov/SCIENCE/Volatiles/

\noindent
Porcello, L.J, et al.\ 1974, Proc.\ IEEE, 62, 769.

\noindent
Ramanan, R.V.\ \& Adimurthy, V.\ 2005, J.\ Earth Syst.\ Sci., Dec. 619.

\noindent
Rivkin, A.S., Howell, E.S., Britt, D.T., Lebofsky, L.A., Nolan, M.C.\ Branston,
D.D.\ 1995, Icarus, 117, 90

\noindent
Rivkin et al. 2002, Asteroids III, 237

\noindent
Ross, S.D.\ 2006, Am.\ Sci., 94, 230.

\noindent
Saal, A.E., Hauri, E.H., Rutherford, M.J.\ \& Cooper, R.F.\ 2007,
Lun.\ Plan.\ Sci.\ Conf., 38, 2148.

\noindent
Simpson, R.A.\ 1998, in ``Workshop on New Views of the Moon,''
eds.\ B.L.\ Jolliff \& G.\ Ryder (LPI: Houston), p.\ 61.

\noindent
Stacy, N.J.S.\ 1993, Ph.D.\ thesis (Cornell U.).

\noindent
Tapley, B.J., et al.\ 2005, J.\ Geodesy, 79, 467.

\noindent
Thompson, T.W.\ \& Campbell, B.A.\ 2005, Lun.\ Plan.\ Sci.\ Conf., 36, 1535.

\noindent
Tomanry, A.B.\ and Crotts, A.P.S.\ 1996, AJ, 112, 2872.

\noindent
Tubbs, R.N.\ 2003, Ph.D.\ thesis (University of Cambridge).

\noindent
Vondrak, R.R., Freeman, J.W.\ \& Lindeman, R.A.\ 1974, Lun.\ Plan.\ Sci.\ Conf.,
5, 2945.

\noindent
Watkins, M., Folkner, W.M., Nerem, R.S.\ \& Tapley, B.D.\ 2006, in {\it 
Proc.\ GRACE Science Meeting, 2006 Dec.\ 8-9}, in press
(http://www.csr.utexas.edu/grace/GSTM/2006/a1.html).

\newpage
Table 1: Summary of Basic Experimental/Observational Techniques Detailed Here

\begin{verbatim}
   All methods are Earth-based remote sensing unless specified otherwise.
--------------------------------------------------------------------------------
Goal           Detection Method         Channel  Advantages      Difficulties
-------------  -----------------------  -------  --------------  ---------------

Map of TLP     Imaging monitor, entire  optical  schedulability  nearside only,
activity       nearside, ~2 km resol.            comprehensive;  limited resol.
                                                 more sensitive
                                                 than human eye

Polarimetric   Compare reflectivity in  optical  easy to         requires use
study of dust  two monitors with                 schedule;       two monitors
               perpendicular polarizers          further limits
                                                 dust behavior

Changes in     Adaptive optic imaging,  0.95     "on demand"     undemonstrated,
small, active  ~100 m resolution        micron,  given good      depends on
areas                                   etc.     conditions      seeing; covers
                                                                 ~50 km dia. max

               "Lucky Imaging,"         0.95     on demand       low duty cycle,
               ~200 m resolution        micron,  given good      depends on
                                        etc.     conditions      seeing

               Hubble Space Telescope,  0.95     on demand       currently  
               ~100 m resolution        micron,  given advance   unavailable; 
                                        etc.     notice          low efficiency

               Clementine/LRO/          0.95     existing or     limited epochs;
               Chandrayaan-1 imaging,   micron,  planned survey  low flexibility
               ~100 m resolution        etc.

               SELENE/Chang'e-1         0.95     existing or     limited epochs;
               imaging, higher resol.   micron,  planned survey  low flexibility
                                        etc.

TLP spectrum   Scanning spectrometer    NIR,     may be best     requires alert
               map, then spectra taken  optical  method to find  from TLP image
               during TLP event                  composition &   monitor; limit
                                                 TLP mechanism   to long events

Regolith       NIR hydration bands      2.9,3.4  directly probe  requires alert
hydration      seen before/after TLP    micron   regolith/water  from monitor,
measurement    in NIR imaging                    chemistry;      flexible
                                                 detect water    scheduling

               Scanning spectrometer    2.9,3.4  directly probe  requires alert
               map, then spectra taken  micron   regolith/water  from monitor,
               soon after TLP                    chemistry;      flexible
                                                 detect water    scheduling

Relationship   Simultaneous monitoring  Rn-222   refute/confirm  optical monitor
between TLPs   for optical TLPs and by  alpha &  TLP/outgassing  only covers
& outgassing   SELENE for Rn-222 alpha  optical  correlation;    nearside; more
               particles                         find gas loci   monitors better

Subsurface     Penetrating radar        ~430MHz  directly find   ice signal is
water ice                                        subsurface ice  easily confused
                                                 with existing   with others
                                                 technique

               Penetrating radar from   ~300MHz  better resol.;  ice signal is
               lunar orbit                       can study       easily confused
                                                 sites of lower  with others;
                                                 activity        more expensive

               Surface radar from       > 1GHz   better resol.;  redundant with
               lunar orbit                       study TLP site  high resol. 
                                                 surface change  imaging?

High resol.    Imagers at/near L1, L2   optical  map TLPs with   expensive, but
TLP activity   covering entire Moon,             greater resol.  could piggyback
map            at 100 m resolution               & sensitivity,  communications
                                                 entire Moon     network

Comprehensive  Two Rn-222 alpha         Rn-222   map outgassing  expensive; even
Rn-222 alpha   detectors in polar       alpha    events at full  better response
particle map   orbits 90 degrees apart           sensitivity     w/ 4 detectors

Comprehensive  Two mass spectrometers   ions &   map outgassing  expensive; even
map of outgas  adjacent polar orbits    neutral  events & find   better w/ more
components                                       composition     spectrometers
________________________________________________________________________________
In situ, surface experiments: we refer the reader to work in preparation by
AEOLUS collaboration.

Abbreviations used:
dia. = diameter, max = maximum, NIR = near infrared, resol. = resolution
________________________________________________________________________________
\end{verbatim}
\newpage

\section {FIGURES}

\begin{figure}
\plotfiddle{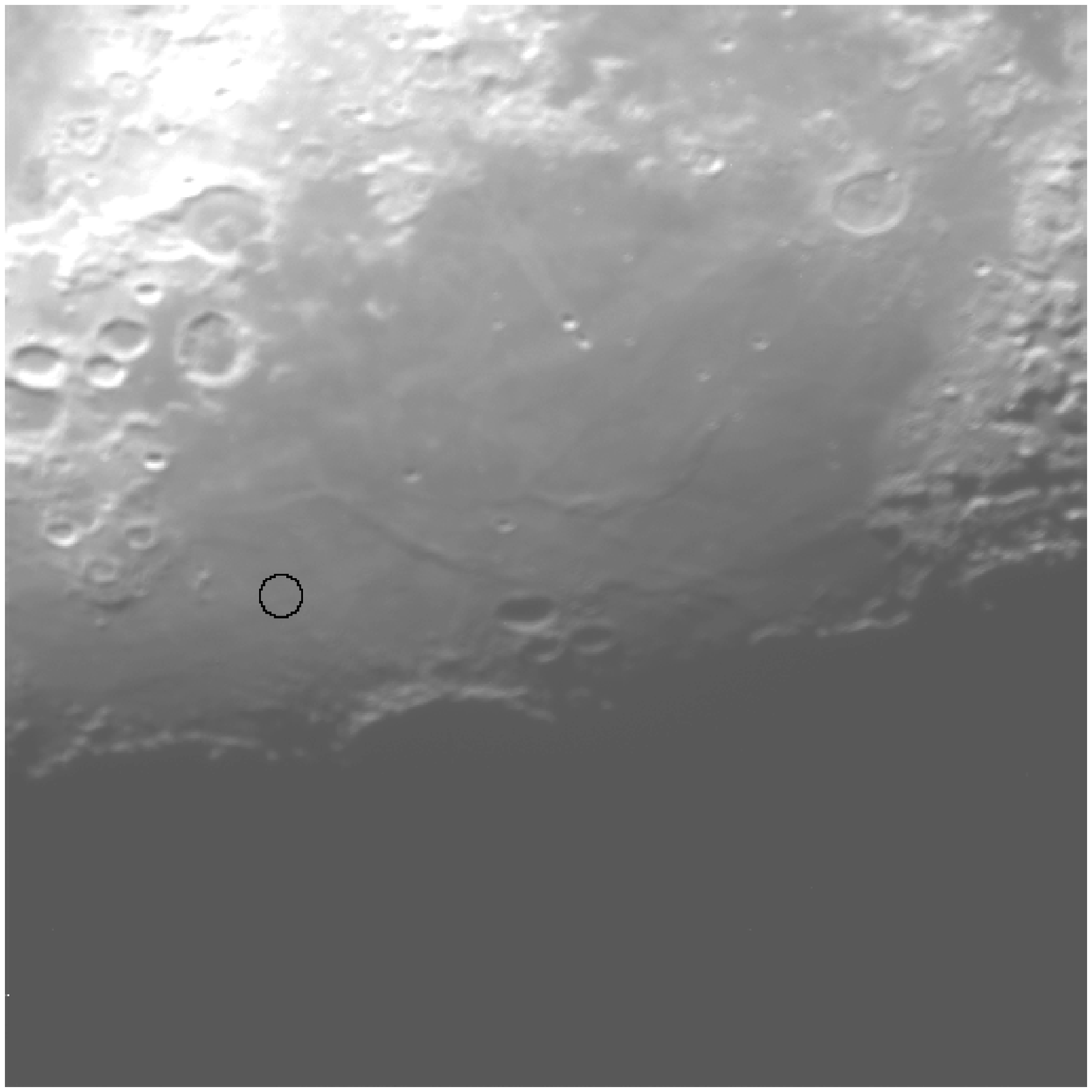}{2.0in}{000}{045}{045}{-127}{-070}
\vskip 0.00in
\caption {
Raw image of a lunar surface subimage typical of what we
expect to achieve with our CCD/telescope combination.
A synthetic signal, corresponding to a TLP below the visual threshold, has
been added at the position marked by the black circle.
}
\end{figure}

\begin{figure}
\plotfiddle{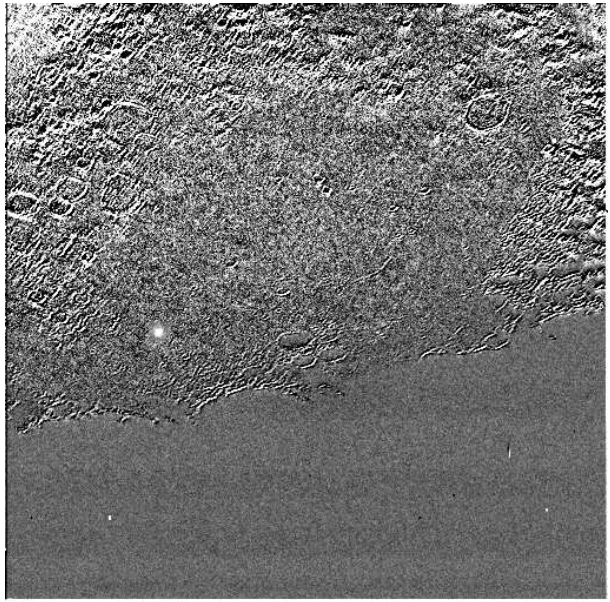}{1.8in}{000}{045}{045}{-127}{-070}
\vskip 0.00in
\caption {
The difference in signal between the image in Figure 2 and
similar one obtained five minutes later.
The noise in the residual signal is at or near the photon limit.
Only the TLP, a few small cosmic rays, and some low-level poor subtraction
residuals in the most complex portion of the image (highland near global
terminator) remain.
}
\end{figure}

\begin{figure}
\plotfiddle{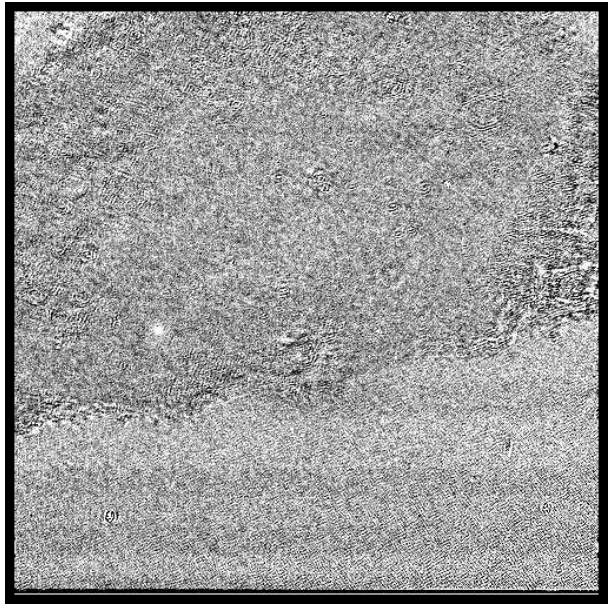}{1.9in}{000}{045}{045}{-127}{-070}
\vskip 0.00in
\caption {
A ``signal-to-noise plot'' of residual signal seen in Figure 2 divided by
the square root of the number of photoelectrons from the signal seen in
Figure 1.
Note that the ``TLP'' stands out above all other signals.
}
\end{figure}

\begin{figure}
\plotfiddle{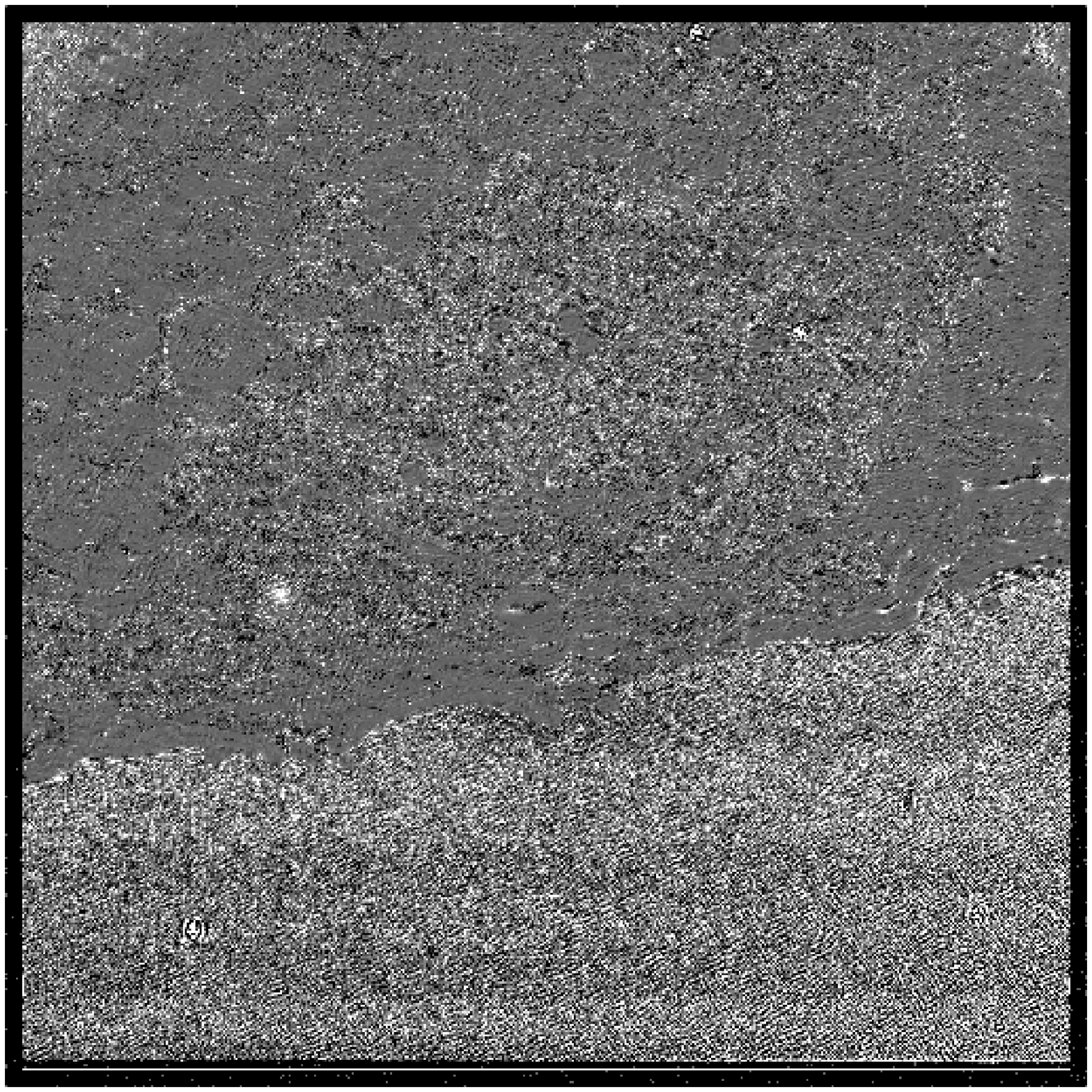}{1.8in}{000}{045}{045}{-127}{-070}
\vskip 0.00in
\caption {
The result from applying a Roberts edge-enhancement filter to 
Figure 1's signal then dividing this into the data from Figure 2.
}
\end{figure}

FIGURES 5 AND 6 ARE LARGE FILES AND INCLUDED AS CAPTIONS ONLY
(see http://www.astro.columbia.edu/~arlin/TLP/ for full Figures 5 and 6.)

FIGURE 5 -
a) Left:
spectrum of an 8-arcmin slit intersecting Aristarchus (bright streak just above
center) and extending over Oceanus Procellarum, and covering wavelengths
5500-10500\AA, taken by the MDM 2.4-meter telescope;
{\bf b)} Right: the residual spectrum once a model consisting of the outer
product the one-dimensional average spectrum from Figure 3a times the
one-dimensional albedo profile from Figure 3a.
The different spectral reflectance of material around Aristarchus is apparent
(at a level of about 7\% of the initial signal), with r.m.s.\ deviations of
about 0.5\%, dominated by interference fringing in the reddest portion, which
can be reduced.

FIGURE 6 -
a) Left: a B-band image of the region around Aristarchus;
b) Right: an image of Aristarchus in a 3\AA-wide centered near 6000\AA,
constructed by taking a vertical slice through Figure 3a and other exposures
from the same sequence of spectra scanning the surface.
Any such band between 5500\AA\ and 10500\AA\ can be constructed in the same
manner, with resolution of about 1km and 3\AA.

%
%
%

\end{document}